\documentclass[prl,nofootinbib,twocolumn,superscriptaddress]{revtex4}
\usepackage{amsmath}
\usepackage{psfrag}
\usepackage{graphicx}

\renewcommand{\O}{{\cal O}}
\newcommand{\mysection}[1]{{\bf #1}}

\newcommand{\5}{{\bf 5}}
\newcommand{\ggut}{SU(5)_{\rm SM}}
\newcommand{\W}{{\cal W}}
\newcommand{\mgut}{M_{\rm GUT}}
\newcommand{\mdm}{m_{\rm DM}}
\newcommand{\mgauge}{{m_{\gamma_d}}}
\newcommand{\mgaugino}{{\tilde m_{\gamma_d}}}
\newcommand{\gauge}{\gamma_d}
\newcommand{\gaugino}{\tilde \gamma_d}
\newcommand{\gravitino}{\tilde G}
\newcommand{\kahler}{{K\"ahler} }

\begin{document}

 \title{Searching for Smoking Gun Signatures of Decaying Dark Matter}
  
 \author{Joshua T. Ruderman} \address{Department of Physics, Princeton University, Princeton,  NJ 08544}
   
 \author{Tomer Volansky} \address{School of Natural Sciences, Institute
   for Advanced Study, Princeton, NJ 08540}

 \begin{abstract}
   Clear methods to differentiate between decaying and annihilating
   dark matter (DM) scenarios are still by and large unavailable. In
   this note, we study the potential astrophysical signatures of a new
   class of hidden sector decaying DM models, which can address the
   recent cosmic ray measurements.  Such models may produce primary
   photons and/or neutrinos at large rates, correlated with the
   leptonic production.  The photon and neutrino spectra will then
   contain sharp features at the TeV scale.  We demonstrate the
   discovery potential for upcoming and future measurements by FERMI,
   HESS, AGIS and IceCube/DeepCore.  We show that these models may be
   discovered in the near future.  Specifically, measurements of
   diffuse gamma rays by FERMI can detect the start of a hard photon
   feature.  We argue that these hard spectra can be produced by
   decaying dark matter and be consistent with current constraints,
   but are difficult to reconcile with models of annihilating DM.
   Consequently the measurement of a hard spectral feature, in correlation with
   the current cosmic ray measurements, will strongly favor decaying
   DM models.  Finally we comment on the preliminary results from the
   Inner Galaxy presented by the FERMI collaboration.
  \end{abstract}
  
  \vskip .2in
  
\maketitle
\mysection{Introduction.}  Recent astrophysical
measurements~\cite{Adriani:2008zr} may constitute indirect detection
of dark matter (DM), the validation of which still awaits unambiguous
evidence.  Such evidence may take the form of associated signals in
photon and neutrino channels, which accompany the leptonic excess.  If
produced from DM decays or annihilations, the energetic leptons imply
that photons are produced from Bremsstrahlung radiation (FSR) and
Inverse Compton Scattering (ICS) off starlight, dust, and the CMB\@.
There can also be neutrinos or photons from lepton or pion decays.
Even if these associated signals are discovered, the nature of dark
matter will remain, in part, obscured.  In particular, it will be hard
to determine whether an excess arises from decays or annihilations of
dark matter.  The reason for this is that the decay and annihilation fluxes
scale differently with the density.  Consequently, the flux differs
between the two scenarios mainly at the Galactic Center (GC), where
the density is highest, but also where the background dominates.

A more promising situation may arise if sharp spectral features are
measured.  Such features have not been conclusively observed in the
electron spectra, but may appear in photon or neutrino measurements.
For the remainder of this letter, we loosely define the hard photons or neutrinos that contribute to a sharp spectral feature, as `primary'.
If discovered, this feature would be smoking gun evidence for DM\@.  
Here we point out that such a signal, if found to be correlated with
the leptonic excess, also has the potential to differentiate between
the decaying and annihilating scenarios.  Indeed, annihilating DM
models that address the leptonic excess and have a sizable branching fraction
into primary photons or neutrinos, are already excluded assuming the DM profiles supported by
current N-body simulations \cite{Navarro:2008kc}.  Conversely, in the decaying
case, the photon or neutrino signals lie just below the sensitivity
reach of current experiments.  As we show, these may be measured in
the near future thereby not only confirming the presence of DM, but
also supporting the decaying DM scenario.

The interpretation of the cosmic ray data requires the
DM to have mass $\sim$ few TeV\@.  Many models have been put forth to
explain the data (see~\cite{Meade:2009iu} and refs. therein) nearly all of
which bifurcate into annihilating and decaying DM scenarios.  In order
to explain the measurements, models of annihilating DM require a surprisingly
large cross-section, three orders of magnitude larger than that of a thermal
weakly interacting massive particle (WIMP).  Such an enhancement can,
in principle, be achieved at low velocities, for example, by the
Sommerfeld effect~\cite{Sommerfeld}, without altering the compelling
features of the WIMP scenario.  In practice, however, it may be
difficult to obtain a large enough enhancement and even if achieved,
annihilating DM models are in tension with various $\gamma$-ray and
neutrino bounds~\cite{Meade:2009iu,Mardon:2009rc}, and constraints
from the recombination epoch~\cite{Slatyer:2009yq}.

The alternative of decaying DM solely replaces the need for the
Sommerfeld enhancement.  If dimension six operators are generated at
or around the GUT scale~\cite{Eichler:1989br}, the decay of the DM
particle naturally explains the observed signals, again without
altering the predictions of the thermal WIMP scenario.  In this case, the
various constraints from $\gamma$-rays, neutrinos and the like are
significantly weaker and such signals may be probed by a variety of experiments~\cite{Arvanitaki:2008hq}.
The drawback of many decaying DM models lies in the
difficulty to naturally explain the lack of hadronic activity, without
introducing significant fine-tuning or complicated and ad hoc
structure.  We present a new class of models where weak-scale DM
decays into a hidden `dark' sector which is broken at the GeV scale
and which communicates with the SM through gauge kinetic
mixing~\cite{RV}.  The lack of hadronic activity is explained through
kinematics, much as in the model of~\cite{ArkaniHamed:2008qn}.

Below we study one representative example of this class of models and
show that DM may naturally decay into primary photons and/or neutrinos
with large branching fractions.  We study the current and prospective
measurements of the spectra at FERMI, HESS, AGIS, SuperK and
IceCube/DeepCore.  Our best-fit of the model to the cosmic ray
measurements is consistent with current constraints and within the
sensitivity reach of the above experiments. We comment on the recent
preliminary FERMI results in the diffuse photon spectrum which are
consistent with the predictions of our model.

\mysection{A Model.}  Let us briefly describe a natural model of
decaying DM which predicts correlated high-energy gamma rays and
neutrinos.  To avoid overproduction of antiprotons, the DM must
predominantly decay into light leptons.  This can be achieved either
through additional symmetries that forbid couplings to hadrons or
through kinematical constraints.  The former is more difficult to
accomplish, especially for decaying DM where the symmetries must be
present at the GUT scale, and typically requires some amount of
fine-tuning.  We concentrate on the case where weak-scale DM decays
into a light state that subsequently decays into leptons, with
antiprotons suppressed by kinematics \cite{MardonComment}.  The DM
itself may or may not be charged under the SM gauge group.  Here we
take DM to be SM-charged, which results in a primary neutrino signal.
Below we present a brief description of the model and refer the reader
to~\cite{RV} for a more comprehensive analysis of this and similar
decaying DM models.

We work within the context of supersymmetric GUTs.  The DM states,
$\chi + \bar\chi $, are taken to be two chiral superfields with
charges $(\5,0) +(\bar \5,0)$ under $\ggut \times U(1)_d$ where $\ggut
\supset SU(3)_C\times SU(2)_W\times U(1)_Y$ is the GUT gauge group.
The predicted signatures to be discussed below do not depend on the
choice of GUT or dark gauge group.  The $U(1)_d$ mixes with the SM
through kinetic mixing,
 \begin{eqnarray}
  \label{eq:3}
  -\frac{\epsilon}{2} \int d^2\theta \ \W_d\W_Y. 
\end{eqnarray}
Here $\W_d(\W_Y)$ is the dark(hypercharge) field strength.  This term can be
generated by heavy bifundamental fields and $\epsilon$ is naturally of
order $\epsilon \sim 10^{-3} - 10^{-4}$ \cite{Dienes:1996zr,ArkaniHamed:2008qn}.  The above gauge mixing induces 
an  effective
Fayet-Iliopoulos term proportional to $\epsilon \langle
D_Y\rangle$, naturally triggering the spontaneous breaking of the
$U(1)_d$ at the GeV scale~\cite{Baumgart:2009tn}.  If this is the
dominant term for the breaking, the dark sector is, to a good
approximation, supersymmetric.  Conversely, the $U(1)_d$ breaking can be
triggered by SUSY-breaking effects  communicated through TeV scale fields charged
under the dark sector~\cite{ArkaniHamed:2008qn}.   In this case the
dark spectrum is not supersymmetric and the dark gaugino can be
heavier or lighter than the dark gauge boson.  

 A dimension six operator of the form,
 \begin{eqnarray}
   \label{eq:2}
   \frac{\alpha_d}{4\pi\mgut^2}\int d^2\theta\ \chi \,  \bar \5_f\W_d^2
 \end{eqnarray}
 induces the DM decay.    Here $\alpha_d=g_d^2/4\pi$ is the dark
 gauge coupling and $\bar\5_f$ denotes a SM multiplet.  Such an
 operator is generated by integrating out
 GUT-scale fields, $X,Y$~\cite{Arvanitaki:2008hq,RV},
 \begin{eqnarray}
   \label{eq:1}
   W = (\mgut + X) Y\bar Y + \mgut X\bar X + \bar X \chi \, \bar \5_f. 
 \end{eqnarray}
 where $X$ and $\bar X$ are singlets and $Y+ \bar Y$ have charges
 $({\bf 1},1) + ({\bf 1},-1)$ respectively.
There are other possible dimension six operators that can decay DM into the dark sector, and the choice of decay operator does not alter the qualitative conclusions of this work.  The decay operator of Eq.~\eqref{eq:1} induces DM decay with a lifetime of
order,
\begin{eqnarray}
  \label{eq:4}
  \tau \simeq 10^{26}~\textrm{sec} \left(\frac{\alpha_d}{30^{-1}}\right)^{-2}\left(\frac{\mdm}{3 \textrm{
        TeV}}\right)^{-5}\left(\frac{\mgut}{10^{15}\textrm{ GeV}}\right)^4
  \end{eqnarray}
 which is the right time scale to explain the astrophysical
 anomalies.    We stress, however, that the DM lifetime is a free
 parameter of the theory.  

 As a consequence of Eq.~\eqref{eq:2}, DM decays into a neutrino
 or sneutrino and two dark gauge bosons or gauginos.  The production
 of primary neutrinos is a generic consequence of DM being
 electrically neutral while carrying hypercharge~\cite{ThankNima}.  The other
 possibility of decaying into the neutral Higgs is disfavored due to
 the antiproton bound from PAMELA\@.  The subsequent decay of the
 sneutrino depends on the MSSM spectrum.  Here we assume that the
 sneutrino is the (SM) NLSP and can decay into a gravitino and a neutrino.
 If the sneutrino is heavier, its decay may produce antiprotons.  The resulting antiproton spectrum,
 however, is expected to
 be rather soft due to the cascade decay.  The
 production of the dark gauge bosons and gauginos is followed by their
decay through the kinetic mixing.  The gauge bosons decay into light SM leptons, $\gamma_d\rightarrow l^+l^-$.  To avoid the
 tension with the antiproton and gamma-ray bounds from the GC, 
 the gauge boson mass, $\mgauge$, is required to be
 sufficiently small, below $\sim$ GeV\@.  The gauginos, on the other
 hand, decay to either dark gauge bosons or photons, as we describe in
 the next section.

 There are several constraints on the model described above.  If DM
 couples elastically to the $Z$, the model is ruled out by direct
 detection \cite{Ahmed:2008eu} by 2-3 orders of magnitude
 \cite{Cirelli:2005uq}.  This constraint can be evaded by introducing
 a dark matter splitting, $\delta m_{\rm DM} \gtrsim 100$ keV, such
 that DM couples inelastically to the $Z$.  With a splitting, there is
 a strong constraint from neutrino telescopes on inelastic capture in
 the Sun, followed by annihilation into $Z Z$ or $W^+ W^-$
 \cite{Nussinov:2009ft}.  This constraint can be evaded by taking a
 slightly larger splitting of $\delta m_{\rm DM} \gtrsim 500$ keV\@.  We
 evade these constraints by coupling DM to the SM Higgs:
\begin{equation}
\label{eq:split}
W_{\rm split} = m_N N^2 +m_{\rm DM} \chi_2 \bar \chi_2 + \chi_2 H_d N
\end{equation}
where $N$ is a singlet with weak-scale mass $M_N$, and $\chi_2$ denotes the doublet component of $\chi$, where the DM resides.  DM stability
requires $M_N > M_{\rm DM}$, and integrating out $N$ generates the
required DM splitting.

 If the dominant DM annihilation channel is to SM gauge bosons,
 its mass is constrained to be of order $1.1$ TeV, in order to obtain the correct thermal abundance~\cite{Cirelli:2005uq}.  Such a low mass is
 inconsistent with the FERMI and HESS measurements.  Interestingly,
 the same operator that generates the splittings to evade the direct
 detection bounds, Eq.~\eqref{eq:split}, also opens a new annihilation channels into Higgses which can
 easily dominate the DM annihilation cross-section.  Consequently, the mass of the DM is a free parameter.

Another important constraint on this model is related to the lifetime of the
triplet partner of the DM, $\chi_3$.  There are strong constraints
on colored particles with $\tau_3 \gtrsim 10^{17}$~sec because
they form exotic atoms \cite{Yamagata:1993jq}.  These constraints are
evaded if the triplet partner decays through a dimension-5 operator,
which we take to be in the \kahler potential,
\begin{equation}
\frac{1}{\mgut} \int d^4 \theta \  \chi \, {\bf \bar 5}_f^\dagger s\ ,
\end{equation}
where $s$ is a singlet.  This operator decays $\chi_3$ in
$\tau_3 \sim 1$~sec, but keeps the  DM stable  as long as $m_{\rm DM} < m_s <
m_{\rm \chi_3}$.  That $m_{\rm \chi_3}$ is heavier than $m_{\rm DM}$ is a generic consequence of RG evolution.

 We note that there are a number of operators that must not be
 present, such as dimension-5 operators which induce prompt DM decays,
 and tree-level Yukawa couplings
 between DM and SM matter.  All such operators are easily forbidden by
 symmetries at the GUT scale.  Cosmology also places nontrivial and interesting
 constraints on the spectrum and lifetimes of light fields in the dark
 sector.  For a detailed discussion of both GUT-scale symmetries and
 cosmological constraints on this model, we refer the reader to~\cite{RV}.  See also~\cite{Finkbeiner:2009mi} for discussions of the cosmology of light hidden sectors.

 \begin{figure*}[t] 
 \includegraphics[width=7.3in]{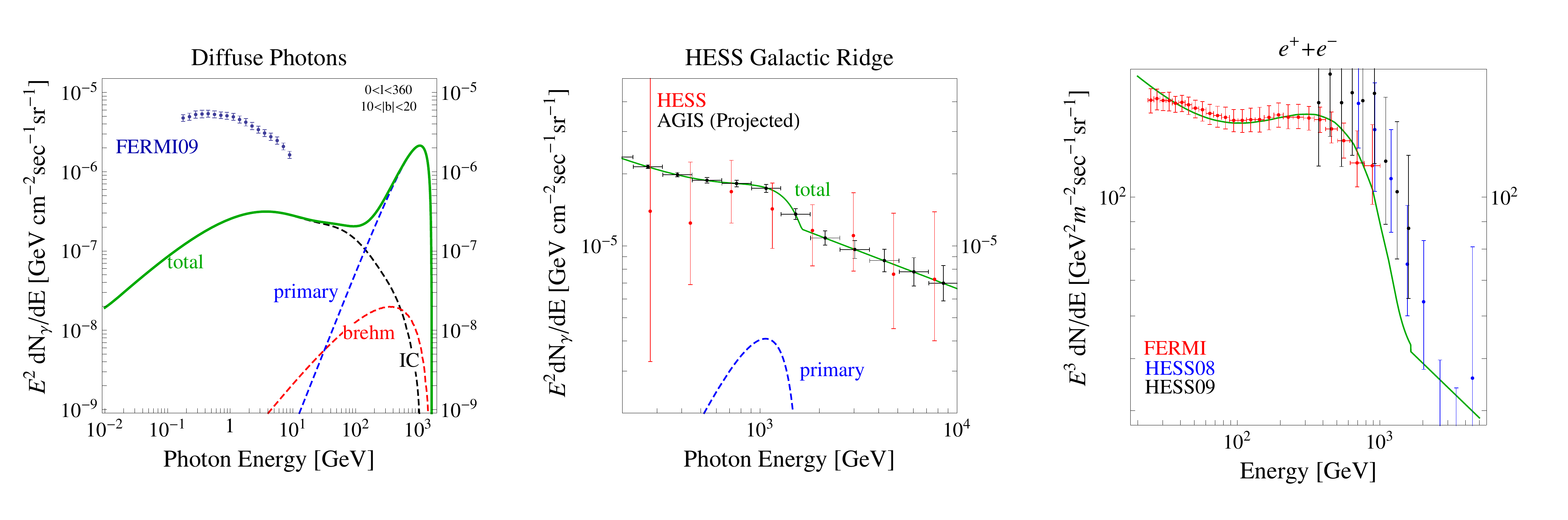} 
 \caption{The electron fit and photon flux predictions for our best-fit model, using the Einasto DM profile with $\alpha=0.17$.
   The {\it right} plot shows the fitted curve to the $e^++e^-$
   flux measured by FERMI and HESS~\cite{Adriani:2008zr}.  The {\it middle} curve shows the predicted
   spectrum from the Galactic Ridge.  The red data points represent the
   corresponding HESS measurement~\cite{Aharonian:2006wh} with $2\sigma$ error bars.  The dashed blue line is the primary photon component predicted by our model, the FSR contribution is too small to be plotted,
  and the green line is the total flux, including a background modeled by a power-law
   and fitted to the HESS data.  The signal is consistent with the current HESS data and can be constrained by further statistics.   The
   black data points illustrate the projected reach for the AGIS
   experiment~\cite{AGIS} with $2\sigma$ error bars.  As can be seen, AGIS will decisively test this scenario.  The {\it left} plot shows the diffuse
   photon prediction in the $0^\circ<l<360^\circ$,
   $10^\circ<|b|<20^\circ$ region.  The blue points correspond to the
   recent FERMI measurement~\cite{Porter:2009sg}.  The dashed lines show the primary (blue), ICS (black),
   and FSR (red) contributions to the total flux, shown in green.  As the FERMI data extends to higher energies it may be able to measure the turnover between the ICS and primary photon components of the signal.}
   \label{fig:photons}
\end{figure*} 
 
\mysection{Gamma-ray Signatures.}  The measurement of a photon line
or sharp spectral feature is considered to be a smoking gun
signature for DM\@.  Typically, such features are model dependent and
are associated with the high end of the spectrum, set by the DM mass.  Interestingly,
within the context of the observed electronic activity, a measurement
of this kind would point towards the decaying rather than annihilating
DM scenario.  Indeed, as was shown
in~\cite{Meade:2009iu,Mardon:2009rc}, annihilating DM models that
provide explanations of the PAMELA and FERMI anomalies and which
have sizeable branching fractions into primary photons or $\pi^0$'s are
excluded.  The exclusion holds under the assumption of DM profiles
suggested by state-of-the-art N-body simulations.  To
ameliorate it, one would have to resort to currently unsubstantiated profiles (see however \cite{RomanoDiaz:2008wz}) 
which predict a significantly shallower or flatter density, $\rho$, at the
GC.  A key observation is then: \emph{if measurements of photons that
  are correlated with the PAMELA and FERMI results indicate the
  production of primary photons or photons from $\pi^0$'s, then decaying
  DM models are strongly favored.}

Models of DM in which the DM decays or annihilates into a light hidden
sector, predict gamma rays from several channels.  Three irreducible
sources of photons are FSR, ICS and synchrotron
radiation.  Concentrating on the high energy spectrum, the first two
dominate, where FSR dominates at the high end of the spectrum while
the ICS dominates at lower energies.  Two additional sources of
photons are those produced from $\pi^0\rightarrow \gamma\gamma$ decays
and primary photons.  In this letter we concentrate on the latter.  Whether
primary photons are produced depends strongly on the dark spectrum.  Assuming a light gravitino, if
the dark gauginos are heavier than the dark gauge bosons, $\mgaugino >\mgauge$,
the decay $\gaugino\rightarrow \gauge \,\gravitino$ dominates and no primary
photons are produced.  If, on the other hand, the dark gaugino is
degenerate with or lighter than the dark gauge boson, it decays through
the kinetic mixing to a SM photon and a gravitino:
\begin{eqnarray}
  \label{eq:5}
  \gaugino \rightarrow \gamma + \gravitino.
\end{eqnarray}
Measurement of the primary photons produced through this decay is thus an
indication of a light dark gaugino (or more generally a light fermion) in the
dark sector.  Conversely, the absence of such photons indicates a
heavy fermionic spectrum, and in particular implies significant
SUSY breaking in the dark sector.
We stress that as opposed to the primary neutrino signal to be discussed in the
next section, and which follows from the specific operator of Eq.~\eqref{eq:2}, the primary photon signal is rather generic for models where dark matter decays or annihilates into a light hidden sector, and the signal probes the lower
end of the dark sector spectrum.
 
In the model discussed above, the $\chi$ supermultiplet is split, and we
assume that the fermion plays the role of the DM\@.  
We use 3-body phase space spectra with flat matrix elements, which
captures the model-independent part of the spectrum.  The resulting
primary photons arise from the 
two available decay channels:
\begin{eqnarray}
  \label{eq:6}
  \tilde \chi \rightarrow \tilde\nu\, \gaugino\, \gauge\ ,\qquad    \tilde \chi \rightarrow \nu\, \gaugino\,\gaugino\,.
\end{eqnarray}
and the photon spectrum depends on the mass of the sneutrino, which we keep as a free
parameter.  We begin by fitting our model to the PAMELA, FERMI, and
HESS electron and positron cosmic rays spectra, as shown in
Fig.~\ref{fig:photons}c.  We obtain the best-fit values:
$m_\chi~=~3.3$~TeV, $m_{\tilde \nu}~=~370$~GeV,
$m_{\gamma_d}~=~400$~MeV, and $\tau_\chi~=~4 \times 10^{26}$ sec.
Consequently, the dark photon  decays into electrons and  muons with branching
 fractions $(0.75, 0.25)$ respectively.   We use this best fit model for the gamma ray and neutrino analyses that follow.

We compute contributions to the photon flux from the primary photons described above, ICS, and FSR\@.  No $\pi^0$'s are produced due to the lightness of $\gauge$.  To
 compute the ICS we propagate the electrons following the procedure
 described in~\cite{Meade:2009iu}.   Throughout the analysis we assume
 the MED propagation model~\cite{Donato:2003xg} and Einasto DM
 profile~\cite{einasto},
 \begin{eqnarray}
   \label{eq:7}
\rho(r) = \rho_{\odot}\textrm{
  Exp}\left[-\frac{2}{\alpha}\left((r/r_s)^\alpha - 1\right)\right]
 \end{eqnarray}
 with $r_s = 20$ kpc, $\rho_\odot = 0.3$ GeV/cm${}^3$ and $\alpha =
 0.17$.  Since we're studying decaying DM, the dependence on the
 choice of profile is significantly weaker than in the annihilating
 case.  Our gamma ray results are relatively insensitive to varying $\alpha$ in the range found in simulations, $0.12 - 0.2$, and to interchanging Einasto with NFW~\cite{Navarro:1995iw}\@.  The flux does not change by more than a
 factor of 3 around the GC nor does it vary by more than $10\%$ for the
 diffuse gamma signal in the regions studied below.

 We consider signals from both the center of the Galaxy and diffuse
 gammas.  A measurement of the latter in the region $0^\circ< l <
 360^\circ$, $10^\circ < |b| < 20^\circ$ was recently released by the
 FERMI collaboration~\cite{Porter:2009sg}.  The data and the
 predictions of the model are shown in
 Fig.~\ref{fig:photons}a.  The blue dashed line corresponds to the primary
 photon spectrum resulting from the DM decays, the red dashed
 line corresponds to the contribution from FSR and the black corresponds to
 that from ICS\@.   As can be seen, the primary photons dominate
 the high end of the spectrum, starting at $\sim 100$ GeV\@.  Since
 FERMI can measure the spectrum up to $300$ GeV, it may only
 probe the tail of the primary photon contribution.
 Although it will be difficult for FERMI to disentangle the primary
 contribution from the ICS, it is possible that it will have enough
 sensitivity to resolve the turnover in the spectrum.  Finally we note
 that a gauge boson with a mass larger than the $400$ MeV presented
 here, can produce a more significant FSR signal.   For complimentary studies of such signals see~\cite{Bertone:2007aw}.

\begin{figure*}[t] 
 \includegraphics[width=5in]{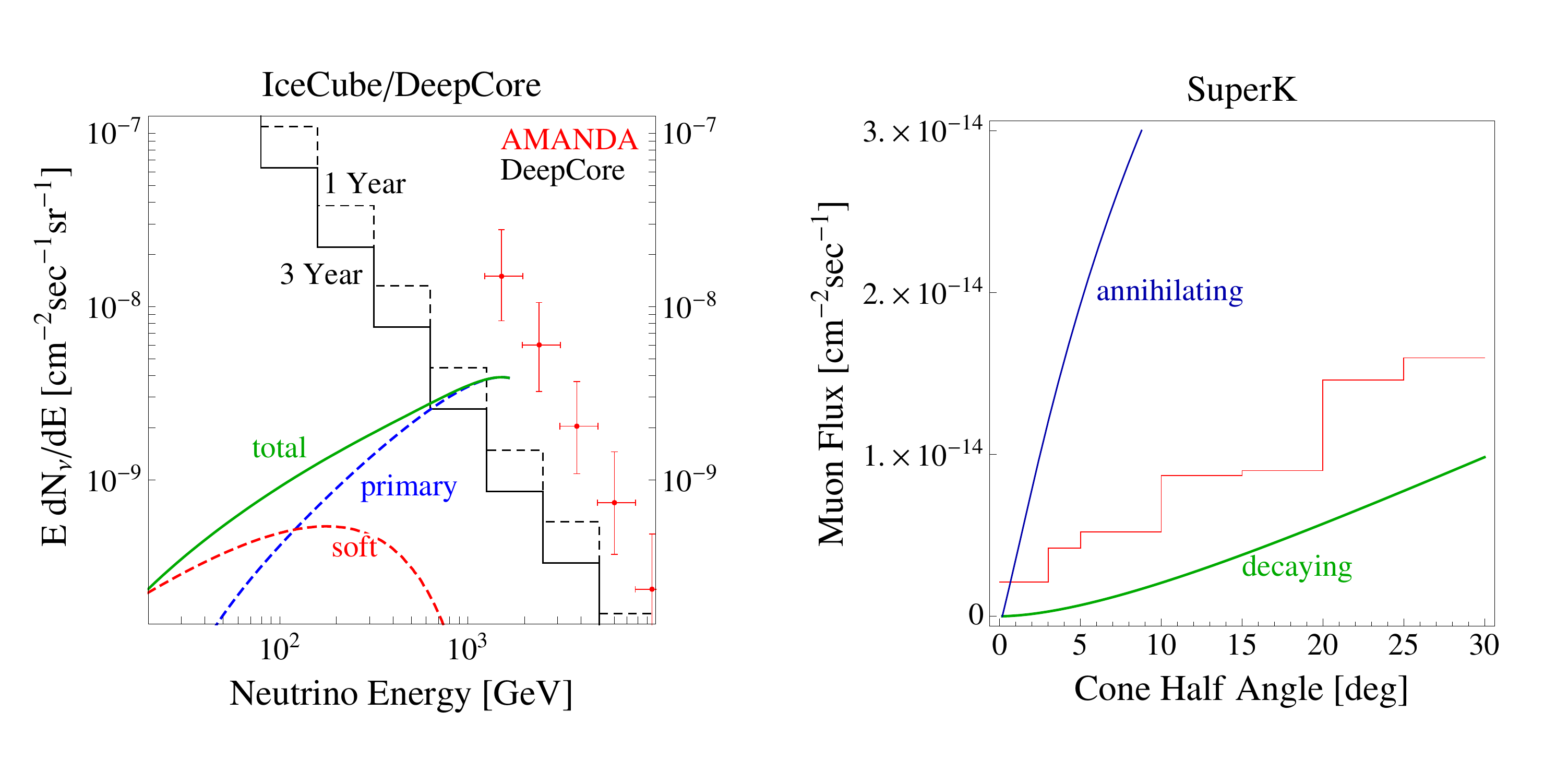} 
 \caption{Neutrino constraints and discovery reach.  The {\it right} plot
   shows the 95\% C.L. muon flux limit from SuperK \cite{Desai:2004pq} as a function of the
   opening angle around the Galactic Center.  The green curve corresponds  to our best-fit
   decaying DM model, while the blue curve illustrates the
   expected flux for an annihilating model with a large branching fraction of primary neutrinos.
   Our decaying model is consistent with the constraints, while the annihilating model is excluded.
     The {\it left}
   plot shows the neutrino flux as a function of the energy.  The
   green curve corresponds to the predicted total signal and the dashed
 lines show the primary (blue) and soft (red) neutrino components.  The data
 correspond to the AMANDA measurement \cite{:2007td}.  The black lines indicate the
 expected 1- (dashed) and 3- (solid) year $5\sigma$ discovery reach in
 IceCube/DeepCore \cite{Resconi:2008fe}.  We find that there will be sufficient sensitivity to detect the flux in the last two bins of the predicted signal.}
   \label{fig:neutrinos}
\end{figure*} 

A remark is in order.  As we mention before, the
measurement of the hard
spectrum is a strong indication of decaying DM\@.  The normalization of
the spectrum is such that if a signal resulting from annihilation is seen in the diffused gamma
region, away from the GC, it must be over-produced at the center itself. 
This is the case, for instance, when DM annihilates into two $\tau$'s
which consequently decays into $\pi^0$'s (see Fig. 8
of~\cite{Meade:2009iu}). This statement is in sharp contrast to the
decaying DM scenario since the photon flux depends on $\rho^2$ in
the former case and on $\rho$ in the latter.  Nevertheless, it is conceivable
that the DM density will turn out to be shallower in which case the GC
signal can be hiding below the background.  A correlated measurement
in the two regions would then imply the existence of decaying DM or
some coincidence regarding the DM density at the GC.

Measurements from the GC and Galactic Ridge (GR) have been taken by the HESS collaboration and already place strong constraints on models of DM~\cite{Aharonian:2006wh}.  The GR is more sensitive
since is measures a  smaller flux in a larger region.  Furthermore, it
is less sensitive to the DM profile since the GC itself
(a $0.1^\circ$ cone around the center) is subtracted.  We show this
measurement and the model prediction in Fig.~\ref{fig:photons}b, where the
error bars correspond to $2\sigma$.    The red dashed
line is the signal and the green line corresponds to the total signal plus
background, where the latter is found by fitting a power-law to the
data.  Our predicted signal is below the current HESS 
resolution.  There is also a preliminary FERMI measurement from the GC~\cite{FermiGC} and we find that the predicted signal of our model is well below the data.

In addition to upcoming HESS results with improved sensitivity, there are two future
Cherenkov telescope arrays under consideration: AGIS~\cite{Buckley:2008ud} and CTA~\cite{CTA}.   Both arrays are expected to be significantly larger
than HESS, with an order 5-10 improvement in sensitivity at the
energies of interest.  To demonstrate the capabilities of these
experiments, we concentrate on AGIS and project a possible future
measurement using an effective area of $1\textrm{ km}^2$, energy
resolution of $15\%$ and $200$ hours of measurement.  We show the
projected measurement on Fig.~\ref{fig:photons}b, in black.   The signal is found to be well above
the expected sensitivity~\cite{AGIS}.   To conclude, we remark that
the future location of AGIS will influence its ability to take
measurements from the GC\@.  It is therefore preferable to locate it in
the southern hemisphere, such as the proposed site near the F\'elix Aguilar Observatory in Argentina. 

\mysection{Neutrino Signatures.}  The second smoking gun signature of
DM is a neutrino line or a sharp spectral feature.  While
SuperKamiokande already places important constraints on DM models, the
upcoming experiments IceCube, Antares and KM3NeT will have the
potential to resolve features in the neutrino spectrum at the TeV scale and therefore
discover DM\@.  Here, by a sharp feature, we mean a feature which
appears in one or a few measured bins.  In our model (or in
corresponding annihilating DM models), we therefore distinguish
between primary neutrinos (which are accompanied by a sharp feature),
produced directly from the decay or annihilation, and soft neutrinos
that result from lepton and pion decays.  Within annihilating DM
scenarios that explain the lepton anomalies, primary neutrinos with a
large branching fraction are already excluded.  On the other hand, as
we show below, the corresponding decaying scenarios are right below the
sensitivity of current experiments and can easily be discovered by
future ones.

As a first step, we must make sure that our model is not already
excluded by the current neutrino bounds from SuperK\@.  In
Fig.~\ref{fig:neutrinos}b, we plot the neutrino flux of our best fit
model as a function of opening angle around the GC, integrated over
energies detectable at SuperK.  We find that it is consistent with the
bounds of~\cite{Desai:2004pq}.  We note that annihilating DM models
with order one branching fractions to primary neutrinos are excluded by
this bound by $\O(5-10)$.  Therefore, the detection of a primary neutrino
signal can favor decaying DM\@.  This is illustrated in the figure
where we plot the same spectrum for the case where the dependence on
the rate depends on the density squared, as in annihilating DM.

SuperK can also specifically identify primary neutrinos because at
energies $E_\nu \gtrsim 1$ TeV, there is a high probability of
inducing electromagnetic showers.  For our best fit model we calculate
the muon flux from neutrinos in a cone of $5^\circ$ around the GC\@.
The number of upward showering events is given by the expression~
\cite{Hisano:2009fb}:
\begin{equation}
N^\mu_{\rm shower} = \int d E_\nu \frac{d \phi_\nu}{d E_\nu} f\left(
  E_\nu \right) \epsilon \left( E_\nu  \right) .
\end{equation}
Here $\phi_\nu$ is the neutrino flux, $f\left( E_\nu \right)$ is the
conversion probability of neutrinos into muons of detectable energy,
and $ \epsilon \left( E_\nu \right)$ is the showering probability
extracted from the analysis of~\cite{Hisano:2009fb}.  The result for our
best fit model is
\begin{equation}
\label{eq:superkflux}
N^{\mu}_{\rm shower} = 1.8 \times 10^{-16}~\textrm{cm}^{-2}~\textrm{s}^{-1}.
\end{equation}
This is smaller than the expected background flux of atmospheric
upward showering neutrinos, $N^{\mu}_{\rm bg} = 3.4 \times
10^{-16}~\textrm{cm}^{-2}~\textrm{s}^{-1}$.  SuperK has done an
analysis on upward showering neutrinos from 1646 days of data, and
they find no events in the above cone~\cite{Desai:2007ra}.  Our best
fit model is therefore not excluded, but interestingly close to the
current sensitivity.

The hard neutrino signal predicted by our model will be tested by
upcoming neutrino experiments, such as IceCube and Antares.  Here we
focus on IceCube with DeepCore, which will have good sensitivity for
TeV-scale neutrinos within the DeepCore fiducial
volume~\cite{Resconi:2008fe}.  The surrounding IceCube strings will be
used to veto down-going muons,
such that the main background is due to atmospheric neutrinos.  The
signal is then expected to become comparable to the background at high
energies, because the neutrino-nucleon scattering cross-section rises
as $\sigma_{\nu N} \sim E^2$ while the atmospheric neutrino background
drops rapidly as $E^{-3}$.  We estimate the future sensitivity of
IceCube with DeepCore by following an approach along the lines
of~\cite{Buckley:2009kw}.  The number of neutrino events to be
observed at IceCube, as a function of neutrino energy is given by the
expression,
\begin{equation}
\frac{dN}{dE} = \frac{d \phi_\nu}{dE} \,\rho_{\rm ice} \, N_A \,
\sigma_{\nu N}(E) \, V(E) \, t_{\rm obs}\ ,
\end{equation}
where $\rho_{\rm ice} = 0.9~\mathrm{g}/\textrm{cm}^3$, $N_A =
6.022\times10^{23}~\textrm{g}^{-1}$, $t_{\rm obs}$ is the run time,
$\sigma_{\nu N}$ is the neutrino-nucleon scattering cross-section, and $V(E)
\approx 0.04~\mathrm{km}^3$ is the effective volume of DeepCore for
neutrino showers, roughly estimated from~\cite{Resconi:2008fe}.

We compare our predicted signal at IceCube and DeepCore with the
atmospheric neutrino background~\cite{Honda:2006qj}.  In order to
estimate the energy resolution of IceCube, we bin the background in
bins of size $\log(E_{\rm max} / E_{\rm min})= 0.3$.  In
Fig.~\ref{fig:neutrinos}a, we plot the AMANDA data~\cite{:2007td}
together with the flux sensitivity for $5\sigma$ discovery in each
bin, collecting events from the entire $2 \pi$ sky for 1 and 3 years
of run time.  We also plot the signal flux for our best-fit model.  A
$5 \sigma$ discovery is found to be possible at the highest bin of the predicted spectrum, above $E_\nu
\gtrsim 1$~TeV, after 1 year,  and in 2 bins after 3 years.   This is
in sharp contrast to the soft neutrinos which populate lower energy bins
and therefore cannot be discovered with this method.  
 We see that IceCube
with DeepCore has the potential to decisively test decaying DM with a
primary neutrino component.  It would be interesting to also estimate the
reach of Antares for this scenario.

\begin{figure}[tb] 
 \includegraphics[width=2.5in]{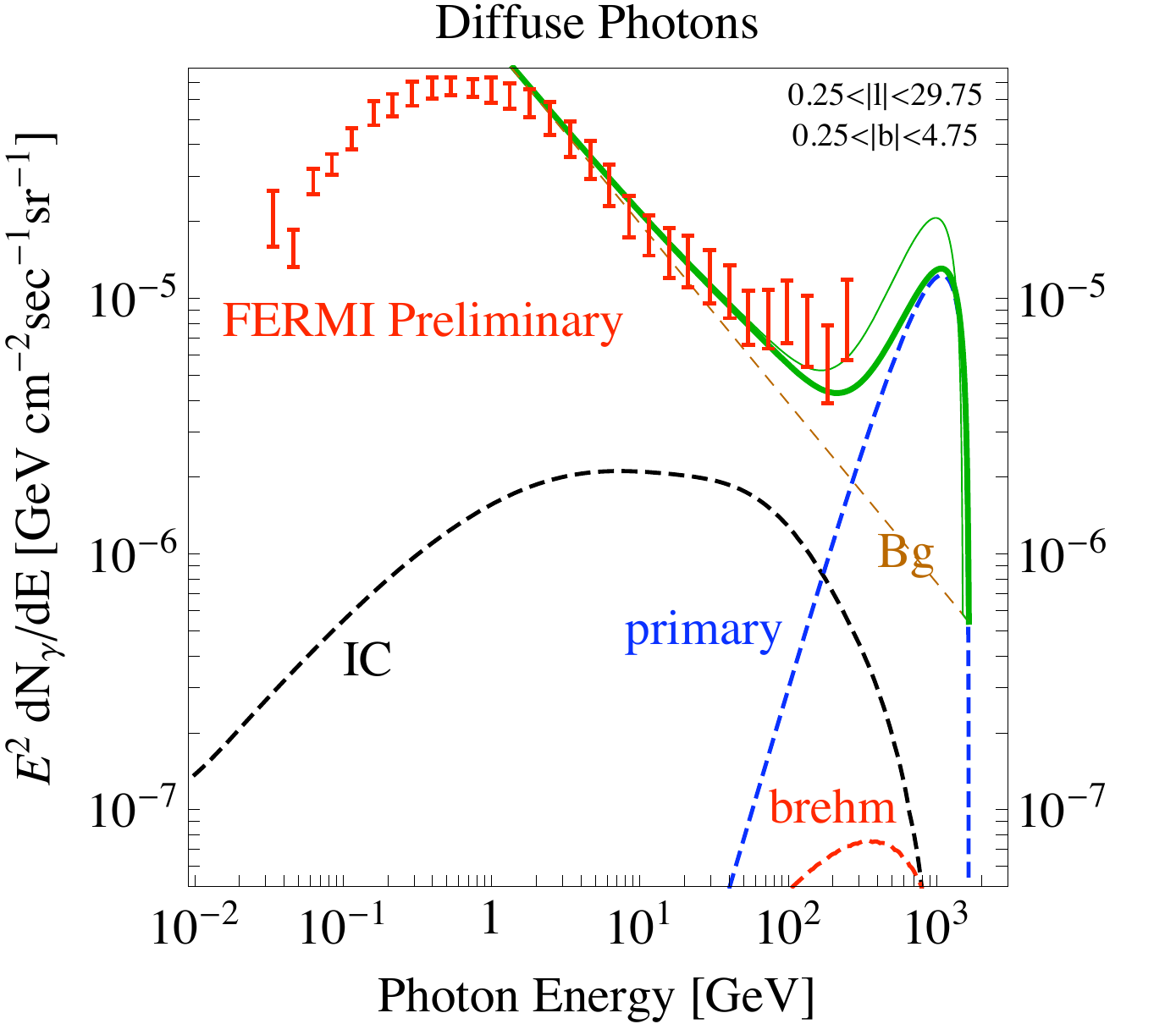} 
 \caption{The prediction of our (i) best-fit model (thick lower green)
 and (ii) lighter DM mass (thin upper green), for the
 diffuse photon spectrum from the Inner Galaxy.  The data correspond to
the preliminary FERMI measurement \cite{FermiIG}.  The background is approximated by
a power-law fitted to the data between $3$ and $16$ GeV\@.  We also plot
the primary, ICS, and FSR contributions to the signal of our best-fit model.  The prediction of our best-fit model is consistent with the data, although it is slightly lower in the last bins.   This is due to slight tension discussed in the text, between the preliminary data and constraints from HESS measurements of Galactic Center photons and the electron cosmic ray spectrum.}
   \label{fig:fermipre}
\end{figure}

\mysection{Note Added.}  While this work was in completion, new but
very preliminary data from the Inner Galaxy region,
$0.25^\circ<|b|<4.75^\circ$, $0.25^\circ <|l|<29.75^\circ$, was
presented by the FERMI collaboration~\cite{FermiIG}.  Interestingly,
the results indicate a deviation from the naive expected background at
high energy.  Being preliminary, these data may well suffer from
significant cosmic-ray contamination, especially at the high end of
the spectrum.  Nonetheless, we cannot resist to compare our
predictions with the measurement, as we do in Fig.~\ref{fig:fermipre}.
In the figure, we plot two sets of curves for two different fits. The
first (thick lower green line) is the best-fit model described above
and the second (thin upper green line) corresponds to a lighter DM mass, 
resulting in a somewhat worse fit to the HESS electronic measurement.  
The ICS, hard and FSR contributions to the spectrum
of the best-fit model are shown in dashed lines.
A careful background analysis is warranted.  Here we simply assume the
dominance of $\pi^0$ decay at energies above $\sim 1$ GeV which can be
approximated by a power-law spectrum.  We therefore model the
background by fitting a power-law to the data between $3$ and $16$
GeV.
 
With the above cautious remarks in mind, we note the following:
\begin{itemize}
\item If confirmed, the excess may indirectly imply the existence of
  DM~\cite{Cholis:2008wq}.
\item For our best fit model, with decay rate normalized to explain
  the PAMELA, FERMI, and HESS leptonic data, there is insufficient ICS
  to account for the excess at high energies $\gtrsim 100$~GeV\@.
  The FSR that we find is also insufficient to account for the excess.
\item For our best fit model, the primary photon component is consistent
  with the preliminary data at high-energies.  Still, our model's
  prediction is slightly below the data.  We find two sources of tension: 
  (i) fitting to the HESS electron measurement results in
  a high DM mass, setting the hardness of the photon spectrum, and
  (ii) the decay rate cannot be increased without introducing 
  tension with the HESS GC/GR measurements.  We expect the overall fit to improve
  should the data points move down following a complete analysis.
\item Future improvement of the statistics may be capable of
  indicating a turnover in the spectrum.  
\item Ref.~\cite{Cholis:2009gv} presents an interesting analysis of
  the data and its implications to models of annihilating DM.  We
  comment that the authors utilize a different fitting procedure from
  the one presented here.  While we fit our decaying model to the
  PAMELA, FERMI and HESS leptonic results, and present predictions for
  the photon and neutrino signals, \cite{Cholis:2009gv} fit only to
  the diffuse gamma excess.  
  The disagreement with regards to the ICS contribution may arise, in part, from different DM masses and different propagation models. 
  It would be interesting to see if
  annihilating models can provide consistent fits to both the leptonic
  and diffuse gamma excesses.
\end{itemize}

\mysection{Acknowledgements.}  We thank N.~Arkani-Hamed, Z.~Komargodski,
M.~Papucci, J.~Thaler, and M.~Vivier for useful
discussions. 
We also thank P.~Meade and N.~Weiner for discussions and comments on the manuscript.  
 We would especially like to thank J.~Buckley and V.~Vassiliev for a very helpful discussion on the future sensitivity of AGIS.  J.~T.~R. is supported by an NSF graduate fellowship.  T.~V. is supported by DOE grant DE-FG02-90ER40542.

\end{document}